# Observation of a Glassy Phase in Solid $^4$He in the Supersolidity Region

V.N.Grigor'ev, V.A.Maidanov, V.Yu.Rubanskii, S.P.Rubets, E.Ya.Rudavskii, A.S.Rybalko, Ye.V.Syrnikov, and V.A.Tikhii

*B.I.Verkin Institute for Low Temperature Physics and Engineering, NAS of Ukraine, 47, Lenin Ave., Kharkov, 61103, Ukraine*
*e-mail: rudavskii@ilt.kharkov.ua*

*High-precision pressure measurements in solid $^4$He, grown by the capillary blocking technique, have been made in temperatures range from 50 to 500 mK. The temperature dependence of pressure indicates that aside from the usual phonon contribution $\sim T^4$, there is an additional contribution $\sim T^2$, the latter becoming dominant at temperatures $T < 300$ mK, where an abnormal behavior attributed to supersolidity has been observed. The data suggest the appearance of a glassy phase (that might be responsible for the anomalous behaviors observed previously).*

## 1. INTRODUCTION

The concept of supersolidity in quantum crystals stated in [1-3] assumed the concurrent existence of crystalline order and superfluidity of some subsystem (for example, vacancies). This interesting idea has immediately attracted the attention of experimentalists but the first experiments on solid $^4$He [4-8] have given no positive results. Recently, Kim and Chan observed non-classical moment of inertia in their torsion experiments in solid $^4$He at $T < 0.2$ K [9]. The effect was interpreted as manifestation of supersolidity, initiating a new tide of interest to the phenomenon. Independent torsion experiments carried out at different laboratories [10-12] confirmed the anomalous behavior of solid $^4$He at $T < 0.2$ K. These experiments also revealed some new peculiarities, namely, a sample history dependence of the effect and a significant influence of crystal annealing that may result even in a complete suppression of the effect. At the same time the search for superfluid flow in solid $^4$He with pressure difference provided no positive result for high-quality crystals [13-14], but the superflow has been registered in the poor quality crystals [14].

Despite numerous theoretical studies [15-25], there is still no unique understanding of the above processes in solid $^4$He. The majority of the authors conclude that supersolidity is impossible in an ideal quantum crystal [15-16, 19-22, 25], and the observed effects are associated with disorder (grain boundaries, etc.) [16,19,21-22,25]. Suggestions of the existence of superfluid [19] and non-superfluid [23-24] glassy states in solid helium, and their possible role in the effects observed have been recently made.

The most typical manifestation of the glassy state is a linear contribution to the temperature dependence of heat capacity. Though the heat capacity of solid $^4$He was studied rather comprehensively, the situation with its temperature dependence at low temperatures remains controversial. Previous experiments [26-28] revealed a linear contribution to heat capacity but the subsequent measurements [29-32] did not find it. Note that the analysis of data of Ref. [32] made in Ref. [23] showed a little linear contribution in heat capacity. The difference between the results was supposed to be due to the qualities of the samples studied. It should be borne in mind that the contribution of the linear term to the heat capacity observed at low temperatures in [26-28] accounts only to several percents of the calorimeter heat capacity, making its reliable determination difficult.

Since linear contribution to heat capacity corresponds to the quadratic-in-temperature contribution to the system pressure, it is convenient to measure pressure to reveal a glassy contribution. Such measurements are by far advantageous because they can be made with a high precision. In that case, there is no contribution equivalent to that of an empty calorimeter to heat capacity. The search for the glassy contribution to the thermodynamics of helium crystals by measuring precisely the pressure in different quality specimens is the goal of the present work.

## 2. EXPERIMENTAL PROCEDURE

The measurements were carried out with the experimental unit previously used to study $^3$He-$^4$He solid solutions and described in detail in [33]. The $^4$He samples in the form of a disc 9 mm in diameter and 1.5 mm high were inside a cooper cell located on the plate of the refrigerator mixing chamber. The cell upper cap made of beryllium bronze acts as a movable electrode of the Straty-Adams capacitive pressure gauge that measured the crystal pressure *in situ* with a ± 3 Pa resolution. Temperature was measured with a $RuO_2$ resistance thermometer calibrated by the crystallization thermometer based on the $^3He$ melting curve.

The samples were prepared with the use of commercial purity $^4He$ by the capillary blocking technique which, as rule, leads to formation of a polycrystalline sample. To reduce the $^3He$ possible content under producing a high pressure with an adsorption gasificator, the first portion of desorbed helium was removed by pumping, leaving behind no more than 10 ppm $^3He$ in the sample. This is evidenced by the absence of phase separation-induced peculiarities of pressure down to 50 mK.

The experimental procedure was as follows. Immediately upon completing the crystallization, the sample was cooled at a maximum possible rate about 30 mK/min down to ~ 1 K to obtain a maximum concentration of defects. Further cooling was made in a step-like manner with a 30 – 100 mK step to temperatures below 100 mK. Following the isothermal ageing for several hours at a minimum temperature, the crystal was heated in a step-like manner. Then the sample was annealed by exposing to a temperature that was by ~ 10 mK below the melting point after which the above procedure was repeated. In some cases the heating was, on purpose, performed up to a temperature at which the annealing does not begin yet, and the sample was recooled to determine the degree of reproducibility of the results. The experiments were carried out on 17 different crystals in a pressure range of 28 to 43 bar.

## 3. ANALYSIS OF THE TEMPERATURE DEPENDENCE OF PRESSURE

Figure 1 shows a typical temperature dependence of relative pressure for a freshly grown crystal during its first cooling (curve 1). To analyze the results obtained, the contributions of different subsystems to the total pressure of solid helium are considered.

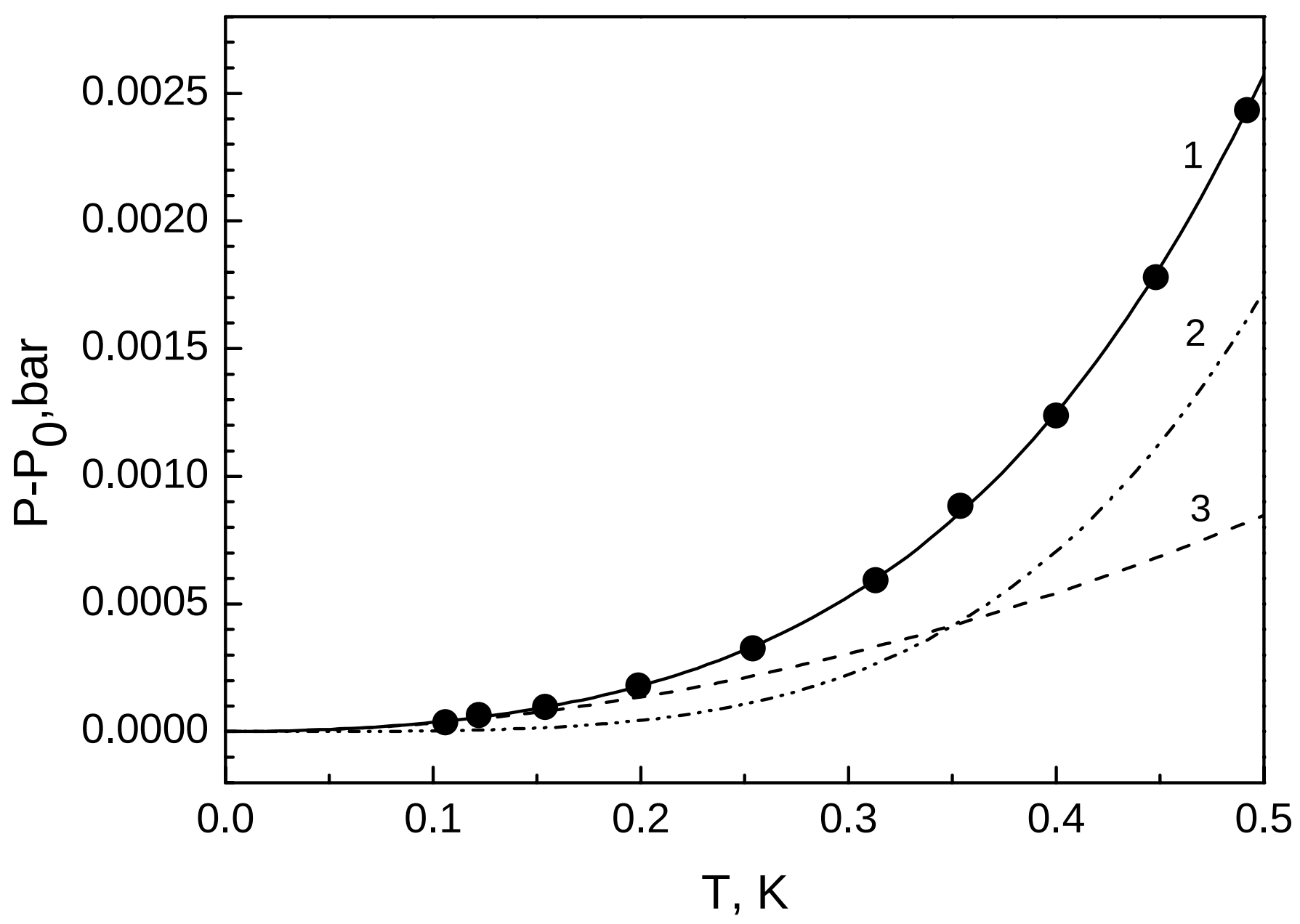

Fig. 1. Typical temperature dependence of relative pressure during cooling the freshly grown sample of solid $^4$He (the solid line 1 is plotted by the experimental points). Curve 2 is the phonon contribution. Curve 3 is glassy contribution (molar volume 20.23 cm$^3$/mol).

As a rule, the temperature dependence of helium crystals is well described by the model in which the total pressure is considered as a sum of phonon, $P_{ph}$(T), and vacancy, $P_{vac}$(T), contributions [34]:

$$P(T) = P_0 + P_{ph}(T) + P_{vac}(T) \tag{1}$$

where $P_0$ is the pressure in the crystal at T = 0. The phonon contribution can be written down as:

$$P_{ph}(T) = \frac{3\pi^4}{5}\frac{\gamma_\Theta}{V}R\frac{T^4}{\Theta_D{}^3} \tag{2}$$

where $\Theta_D$ is the Debye temperature, $\gamma_\Theta = -\frac{\partial \ln \Theta}{\partial \ln V}$ is the Grüneisen constant, R is the gas constant, and V is the crystal molar volume.

The expression for the vacancy contribution to pressure, in terms of the model of wide-band vacancies, is of the form [35]:

$$P_{vac}(T) = R\left(\frac{T}{Q_V}\right)^3 \left(1 + 3\frac{T}{Q_V}\right) \cdot \frac{\gamma_Q}{V} Q_V \cdot \exp\left[-\frac{Q_V}{T}\right] \tag{3}$$

where $Q_v$ is the energy of vacancy formation, $\gamma_Q = -\frac{\partial \ln Q_v}{\partial \ln V}$.

As shown in [34], this model describes well all thermodynamic data for solid $^4$He, $^3$He and $^3$He-$^4$He solutions with taking into account three adjustable parameters $P_o$, $\Theta_D$, and $Q_v$. The analysis yields Debye temperature and activation energy of vacancies, making it possible to obtain universal dependence of these parameters on molar volume for $^4$He, $^3$He and their solutions.

At low temperatures (below ~ 0.5 K) the vacancy contribution to pressure becomes negligible, and the main contribution is made by the phonon subsystem. If one assumes that the abnormal behavior of solid $^4$He observed in various experiments at T < 0.2 K is associated with the formation of a glassy phase, the latter should produce a contribution as $P_g = a_g T^2$ where the coefficient $a_g$ is determined by the density and distribution of tunnel states of glass.

Then the total pressure of the system at low temperature can be given a follows:

$$P(T) = P_0 + P_{ph}(T) + P_g(T) = P_0 + a_p T^4 + a_g T^2 . \tag{4}$$

To emphasize the existence of the quadratic term in the temperature dependence of pressure, Eq. (4) can be rewritten as:

$$(P - P_0)/T^2 = a_g + a_p T^2 \tag{5}$$

Figure 2 shows such dependences as a function of $T^2$ for two samples. The adequate linear dependences suggest the existence of the Debye contribution, and the intersepts on the ordinate axis correspond to the $a_g$ values. However, Eq. (5) is not convenient for quantitative treatment because the experimental points corresponding to different temperatures have nonequivalent accuracy, and introducing weighting factors actually requires Eq. (4). For this reason, Eq. (4) was used in subsequent calculations.

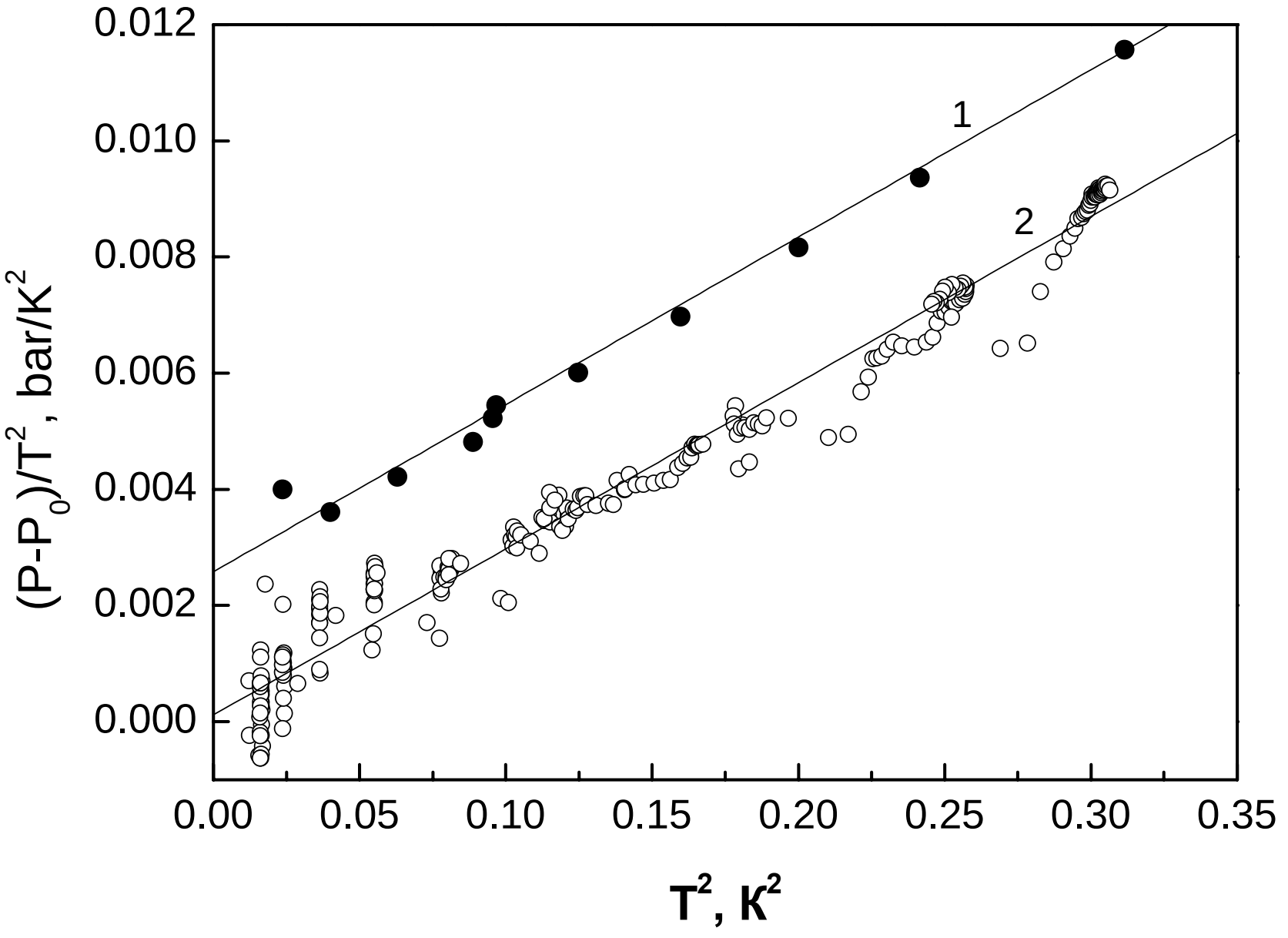

Fig. 2. The $[P(T) - P_0]/T^2$ vs $T^2$ for the freshly grown $^4$He crystal with molar volume 20.23 $cm^3$/mol (curve 1) and annealed crystal with molar volume 19.83 $cm^3$/mol (curve 2).

Since almost all the samples studied were nonequilibrium, the use of the known Debye temperature values for the estimation of phonon contribution is incorrect. So, Eq. (4) contains nominally three adjustable parameters and their determination by rather smooth temperature dependence P(T) becomes fairly uncertain. The problem was somewhat improved because the measurements were carried out down to rather low temperatures at which P(T) was almost independent of $P_0$, and the final version of approximation was chosen from the condition:

$$P(T_{\min}) - P_0 = aT_{\min}^2 \tag{6}$$

where $T_{min}$ is the minimum temperature in the series of measurements where the Debye contribution was negligible.

The values of $a_g$ estimated in this way for not annealed crystals have a wide scatter that does not correlate with the pressure values and ranges from 1.5 to 4 mbar/K$^2$. Such a scatter is quite expectable because the glass contribution is likely to be associated with the concentration of defects which was not controlled. Our value of $a_g$ within the scatter, correspond to linear contribution in heat capacity found in Ref. [28] being however one order of magnitude larger than that of Ref. [32]. The values of coefficient $a_p$ have a much narrower scatter and the Debye temperature values calculated by these values of $a_p$ coincide within 2-3 % with the magnitudes calculated by the universal dependence $\theta_D(V)$ in [34]. However in some cases the deviation was as high as 10%.

The contributions of phonon and glassy components to P(T) are shown in Fig. 1 (curves 2 and 3). As is evident, the glassy contribution becomes dominant at $T < 0.3$ K, i.e. in the temperature range where the "supersolid" effect was observed. Note that recently the Helsinki group observed a small deviation from the phonon $T^4$ dependence at the melting curve below 80 mK [36].

## 4. RELAXATION AND ANNEALING

The experiments on the initially nonequilibrium fresh samples revealed several types of relaxation behaviour. Most of the samples exhibit no considerable relaxation not only on the first cooling but also on the subsequent heating until the temperature approaching the melting point $T_m$ by 100 - 10 mK. The time behaviour of relaxation is characterized by variations in pressure on the samples heating up to $T - T_m \sim 10$ mK is illustrated in Fig. 3 where five typical regions are clearly seen. Region 1 corresponds to the initial state (no relaxation) when the sample was at constant temperature (500 mK) and pressure (43.8 bar). Then the temperature was increased to that of annealing and the pressure first rose with increases in phonon and vacancy contributions with temperature (region 2) and then it decreased when the relaxation contribution was in excess of the influence of temperature rise (region 3). In that case the sharp reduction in pressure at the first stage gave way to a stage of slow relaxation (region 4). And finally, the pressure reached equilibrium in region 5.

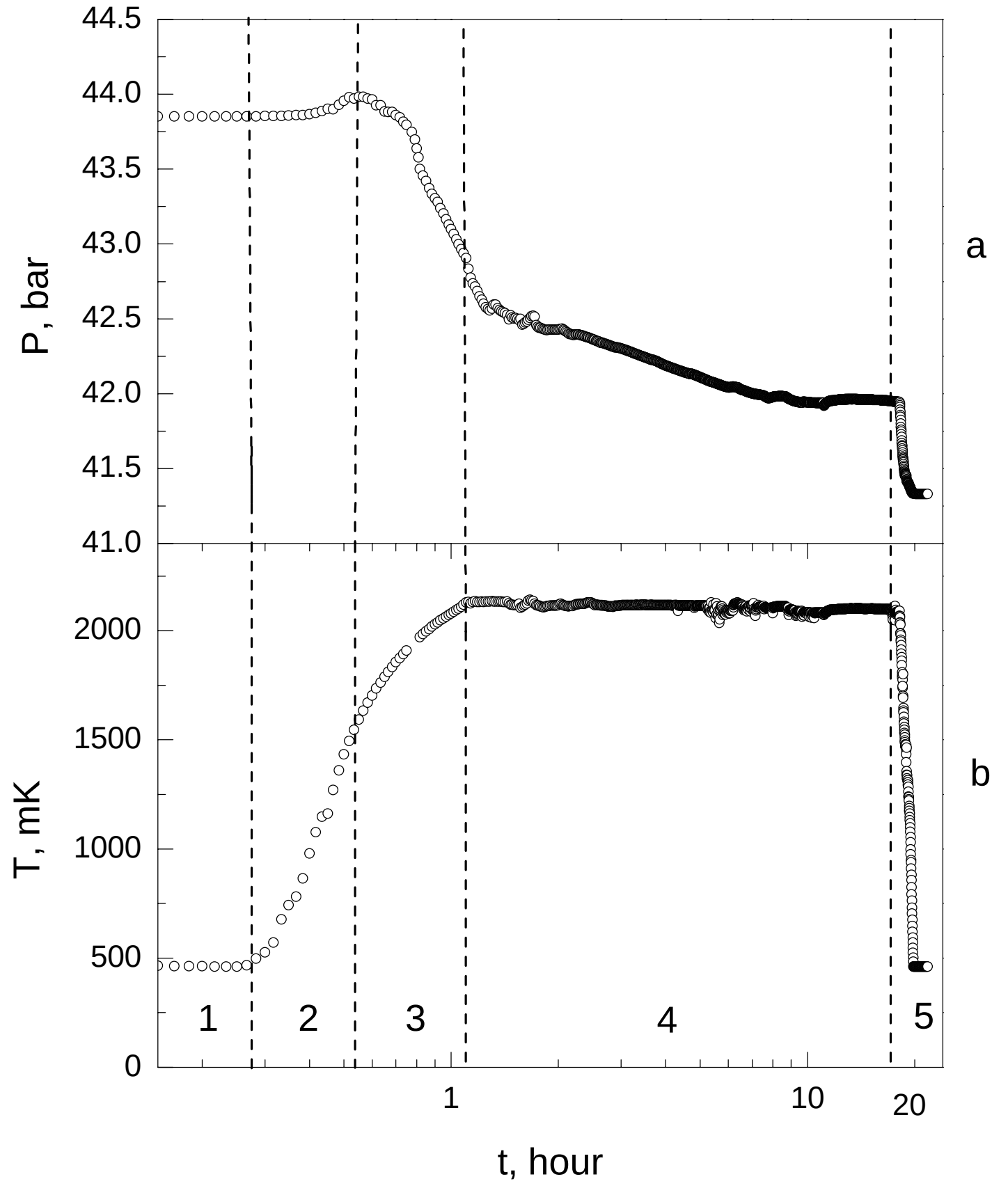

Fig. 3. The kinetics of variation in pressure and temperature of the sample in process of annealing. Initial molar volume 19.61 $cm^3$/mol, final molar volume 19.83 $cm^3$/mol.

Note a considerable difference in pressure, ΔP ~2 bar, before and after annealing at the same temperature. Assuming that the samples studied have an ordinary compressibility of solid helium, this ΔP corresponds to a variation in density by approximately 0.5 %. This effect can hardly be caused by the disappearance of vacancies or dislocations, since it is very unlikely to have that high concentration of such defects. We conjecture that the effect is due to metastable liquid/glass, trapped in "pockets" in the course of fast crystallization. The existence of such pockets on the measuring cell walls was proposed in [37]. Estimating the amount of liquid necessary to quantitatively explain the effect, we get ~ 5 %. This value is too large to be attributed to cell walls, and one comes to the scenario of liquid/glassy cavities in the bulk of the sample. It is tempting to associate the state of $^4$He in these regions with metastable disordered state of quenched $^4$He observed in first-principle Monte Carlo simulations by Boninsegni *et al.* [19]. We also note that the estimated concentration of liquid/glassy inclusions turns out to be close to the superfluid fraction found in Ref. [9].

The data for an annealed sample are presented in Fig. 2 (curve 2). In this experiment, we had no technical possibility to stabilize the temperature with sufficient accuracy. Therefore, the measurements were performed under slow heating and all experimental points were processed In this case the large number of the points compensated their scattering and the experimental error in finding the parameters was practically the same (within 5%) as that under handling the averaged data. The value of $a_g$ was estimated one order of magnitude smaller than in non-annealed crystals. This suggests that annealing of the crystal influences strongly the glass phase contribution.

It should be mentioned that the annealing effect is highly dependent on the duration of annealing region 4 (Fig. 3). If the region duration was one or two hours, the variation in $a_g$ did not exceed the normal scatter. The above two results are consistent with assumption that the main reduction in pressure is caused by the disappearance of "pockets" and that the defects responsible for the glass state disappear at the stage that follows.

In some samples, the pressure relaxation occurred even during the first cooling, sometimes immediately after the onset of cooling and sometimes at subsequent stages. One of the samples displayed no relaxation on cooling but it started on heating up to 300 mK. It should be noted that at each temperature level the pressure was reduced only to a certain value and the next decrease in pressure occurred only with the next increase in temperature. The relaxation process will be discussed in detail elsewhere.

## 5. CONCLUSION

We performed high-precision pressure measurement in solid $^4$He samples grown by capillary blocking technique. In all non-annealed HCP crystals, the temperature dependence of pressure demonstrates a contribution proportional to $T^2$, the latter becomes the leading term at temperatures $T < 300$ mK, at which "supersolid effects" were observed. Such a behavior may be ascribed to a glassy phase. We find that this glassy contribution to pressure can be eliminated only by a substantial annealing.

A dramatic pressure decrease of ~ 2 bar was observed under annealing at temperatures very close to the melting point. We conjecture that this effect is due to solidification of liquid/glass captured in closed cavities during the growth process. To quantitatively account for the pressure drop, the volume fraction of these regions should be as high as ~ 5 %.

The authors are grateful to B .Svistunov and N. Prokof'ev for useful discussions.

The research was supported by CRDF under grant 2853.